\documentclass[11pt]{article}

\usepackage[top=1.0in,right=1.0in,left=1.0in,bottom=1.75in]{geometry}

\usepackage[T1]{fontenc}         
\usepackage[latin1]{inputenc}    
\usepackage{mparhack,fixltx2e,relsize}  
\usepackage{microtype}           
\usepackage{textcomp}            
\usepackage[ngerman,italian,english]{babel}   

\usepackage{amscd,amsmath,amsfonts,amssymb,amsthm}     
\usepackage{bm}                  
\usepackage{braket}              
\usepackage{empheq}              
\usepackage{eucal}               
\usepackage{mathrsfs}            
\usepackage{mathtools}           
\usepackage[output-decimal-marker={.}]{siunitx} 
\usepackage[only,llbracket,rrbracket]{stmaryrd} 

\usepackage{booktabs}            
\usepackage{caption}[2013/02/03] 
\usepackage{graphicx}            
\usepackage{bmpsize}             
\usepackage{multirow}            
\usepackage{rotating}            
\usepackage{subfig}              
\usepackage{tabularx}            
\usepackage[svgnames,dvipsnames,table]{xcolor}  

\usepackage[affil-it,auth-sc]{authblk}           

\usepackage{chngpage,calc,eso-pic,everyshi,ifthen} 
\usepackage[dayofweek]{datetime} 
\usepackage{emptypage}           
\usepackage{afterpage}           
\usepackage{enumerate}
\usepackage{lipsum}              
\usepackage{multicol}            
\usepackage{pdfpages}            
\usepackage[font=small,noorphans]{quoting} 
\usepackage{sectsty}             
\usepackage{setspace}            
\usepackage[nottoc,notlof,notlot]{tocbibind}  
\usepackage{textpos}             
\usepackage{titlesec}            
\usepackage[english]{varioref}   

\usepackage[autostyle,italian=guillemets]{csquotes}   

\usepackage{hyperref}            
\usepackage{bookmark}              

\definecolor{purple}{RGB}{143,52,43}
\def\ds{\displaystyle}
\def\d{\mbox{{d}}}

\usepackage{fancyhdr}            
\fancypagestyle{plain}{%
  \fancyhead{}                                     
  \fancyfoot[CE,CO]{\thepage}       
  \fancyhead[CE,CO]{\textcolor[rgb]{0.64,0.15,0.15}{Published in \emph{Extreme Mechanics Letters} (2015) in press
\\
doi: http://dx.doi.org/10.1016/j.eml.2015.04.007}}
\setlength{\headheight}{14.5pt}}

\begin{document}
\title{Development of configurational forces during the injection of an elastic rod}

\author[1]{F. Bosi}
\author[1]{D. Misseroni}
\author[1]{F. Dal Corso}
\author[1]{D. Bigoni\footnote{Corresponding author: Davide Bigoni, bigoni@unitn.it; +39 0461 282507}}
\affil[1]{DICAM - University of Trento, via Mesiano 77, I-38123 Trento, Italy}

\maketitle

\begin{abstract}
When an inextensible elastic rod is \lq injected' through a sliding sleeve against a fixed constraint, configurational forces are developed, deeply influencing the mechanical response.
This effect, which is a consequence of the change in length of the portion of the rod included between the sliding sleeve
 and the fixed constraint, is
theoretically demonstrated (via integration of the elastica) and experimentally validated on a proof-of-concept structure (displaying an interesting
force reversal in the load/deflection diagram), to provide conclusive evidence to 
mechanical phenomena relevant in several technologies, including guide wire for artery catheterization, or wellbore insertion of a steel pipe.
\end{abstract}


\section{Introduction}

The buckling of a piece of paper hitting an obstacle when ejected from a printer, the insertion of a catheter
in to an artery\cite{Bueck:2004} or of steel piping in to a wellbore\cite{Mitchell:2008, Su-Wicks-Pabon-Bertoldi:2013, Miller:2015, Reis:2015},
even the so-called \lq inverse spaghetti problem' \cite{Carrier:1949, Mansfield:1987, Downer:1993} are all examples of
mechanical settings where an elastic rod is forced out of a constraint and pushed against an obstacle.
During this
process the length of the rod subject to deformation changes, and when the rod is injected\footnote{
Similarly to a fluid in a vein, \lq injection' means that a portion of an elastic rod is forced to enter in a fixed space between two constraints.
}
through a sliding sleeve,
a configurational or Eshelby-like force
\footnote{The concept of configurational force was developed by Eshelby \cite{Eshelby:1951, Eshelby:1956} in the context of solid mechanics to motivate the possibility
of motion of defects, such as dislocations, cracks and voids. In structural mechanics such a force was demonstrated for the first time in \cite{Bigoni-Bosi-DalCorso-Misseroni-2:2014}.}
is generated, that until now has been ignored  in the above-mentioned problems \cite{Ro-Chen-Hong:2010, Ro-Chen:2010}.
This force has  recently been evidenced for a cantilever
beam by Bigoni et al. \cite{Bigoni-Bosi-DalCorso-Misseroni-2:2014}.
It has also been shown to influence stability \cite{Bigoni-Bosi-DalCorso-Misseroni-1:2014}, used
to design a new kind of elastically deformable scale \cite{Bosi-DalCorso-Misseroni-Bigoni:2014} and
to produce a form of torsional locomotion \cite{Bigoni-DalCorso-Misseroni-Bosi-3:2014}.

\begin{figure}[!htcb]
  \begin{center}
     \includegraphics[width= 12 cm]{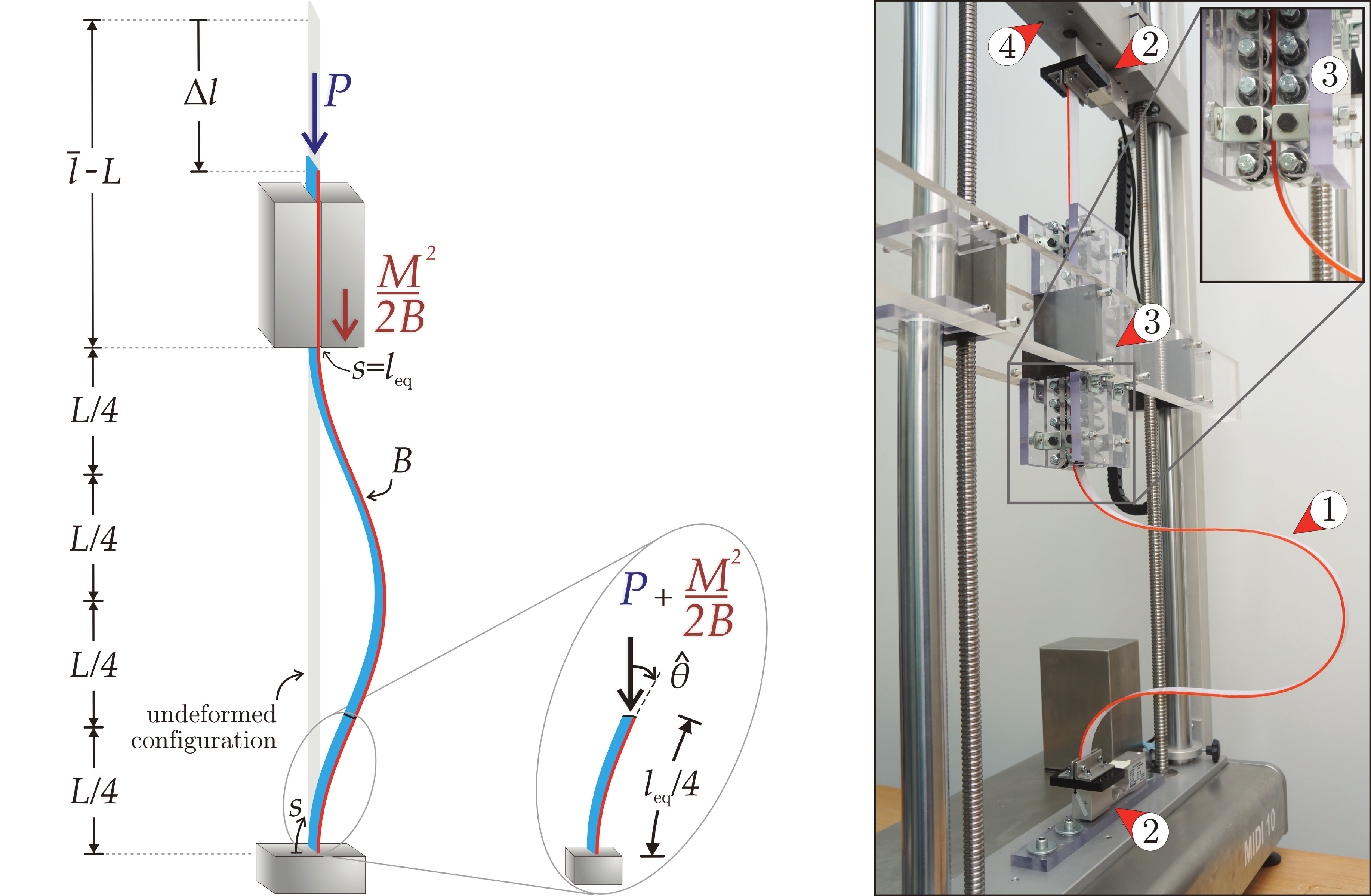}
\caption{\footnotesize Left: Injection of an elastic planar rod through a sliding sleeve via an axial load $P$.
The rod of total length $\bar{l}$ has bending stiffness $B$, is clamped at its lower end
and inserted into a frictionless sliding sleeve (at a distance $L$ from the clamp) at the upper end.
Due to symmetry, the system can be analyzed as four equal cantilever beams of length $l_{\textup{eq}}/4$ loaded with a dead load $P+M^2/2B$, where
$M^2/2B$ is the axial configurational force generated at the sliding sleeve.  Right: the experimental set up realized with an elastic rod (1),
load cells (2), sliding sleeve (3) and movable crosshead (4).}
\label{variablelength}
  \end{center}
\end{figure}

The aim of this article is to provide direct theoretical and experimental evidence that the effect of
configurational forces on the injection of an elastic rod
is dominant and cannot be neglected. 
It leads, in the structure that will be analyzed, to a force reversal that otherwise would not exist.
The considered setup is an inextensible elastic rod, clamped at one end and injected from the other through a sliding sleeve
(via an axial load), so that the rod has to buckle to deflect (Fig. \ref{variablelength}, left).

The elastic system is analytically solved in Section \ref{caricodipunta},
through integration of the elastica (\cite{Love-Libro:1927, Bigoni-Libro:2012}).
It is shown that \lq Eshelby-like' forces strongly influence the loading path and yield a surprising
force reversal, so that certain equilibrium configurations are possible if and only if the applied force changes its sign.
The change of sign is shown to occur when the rotation at the inflexion points exceeds $\pi/2$ (corresponding to $\Delta l/L \approx 1.19$), a purely geometric condition independent
of the bending stiffness $B$ and distance $L$. Furthermore, it is also proven that during loading two points
of the rod come into contact and, starting from this situation (again defined by a purely geometric condition), the subsequent configurations are all self-intersecting elastica.
All theoretical findings have been found to tightly match the
experimental results presented in Section \ref{exppost} and obtained on a structural model
(designed and realized at the Instability Lab of the University of Trento
http://ssmg.unitn.it), see also the supporting electronic material.

\section{The elastica and theoretical behaviour of the structure}\label{caricodipunta}

The structure shown in Fig. \ref{variablelength} (left) is considered axially loaded by the force $P$. The curvilinear coordinate $s \in \left[0,\bar{l}\right]$ is introduced,  where $\bar{l}$ is the total length of the rod, together with the rotation field $\theta(s)$ of the rod's axis. Denoting with a prime the derivative with respect to $s$, the total potential energy of the system can be written as
\begin{equation}\label{eptpost}
\mathcal{V}\left(\theta(s),l_{\textup{out}}\right)=\intop_{0}^{l_{\textup{out}}}B
\frac{\left[\theta^{'}\left(s\right)\right]^{2}}{2}\d s-P\Biggl[\bar{l}-\intop_{0}^{\bar{l}}\cos\theta(s)\d s \Biggr],
\end{equation}
where $l_{\textup{out}}\in[L,\bar{l})$ is the length of the deformed elastic planar rod comprised between
the two constraints (the clamp and the sliding sleeve) and $B$ is the bending stiffness of the rod.

With reference to $\epsilon \ll 1$
(Bigoni et al.~\cite{Bigoni-Bosi-DalCorso-Misseroni-2:2014}) the rotation and the length of the rod comprised between the two constraints
can be written as the sum of the fields evaluated at equilibrium, $\theta_{\textup{eq}}$ and $l_{\textup{eq}}$, and their
variation $\theta_{\textup{var}}$ and $l_{\textup{var}}$
\begin{equation}
\begin{array}{cc}\label{banale}
\theta(s,\epsilon)=\theta_{\textup{eq}}(s)+\epsilon\,\theta_{\textup{var}}(s),\quad\quad
& l_{\textup{out}}(\epsilon)=l_{\textup{eq}}+\epsilon\,l_{\textup{var}},
\end{array}
\end{equation}
so that consideration of the boundary conditions at the sliding sleeve, $\theta_{\textup{eq}}(l_{\textup{eq}})=0$ and $\theta(l_{\textup{out}})=0$, leads to the compatibility equation
\begin{equation}\label{eq:compatcond}
\theta_{\textup{var}}(l_{\textup{eq}})=-\theta_{\textup{eq}}^{'}(l_{\textup{eq}})\,l_{\textup{var}},
\end{equation}
from which the first variation of the functional $\mathcal{V}$ follows as
\begin{equation}
\begin{split}\label{varprima1}
\ds \delta_\epsilon\mathcal{V}&=-\intop_{0}^{l_{\textup{eq}}}\left[B
\theta_{\textup{eq}}^{''}+P\sin\theta_{\textup{eq}}(s)\right] \theta_{\textup{var}}(s)\d s
 - \frac{B}{2}\theta_{\textup{eq}}^{'}(l_{\textup{eq}})^{2}l_{\textup{var}}.
\end{split}
\end{equation}
The fact that the distance $L$ between the two constraints remains constant for every deformation is expressed by the constraint
\begin{equation}
L=\intop_{0}^{l_{\textup{out}}}\cos\theta(s)\d s ,
\end{equation}
so that an additional compatibility condition is obtained
\begin{equation}\label{acaso}
l_{\textup{var}}=\intop_{0}^{l_{\textup{eq}}}\sin\theta_{\textup{eq}}(s)\theta_{\textup{var}}(s) \d s,
\end{equation}
and the vanishing of the first variation $\delta_\epsilon\mathcal{V}$~(\ref{varprima1}) leads to the governing equation of the elastica
\begin{equation}
\label{elasticazzipost}
\ds \theta_{\textup{eq}}^{''}(s) + \left(P+\frac{B}{2}\theta_{\textup{eq}}^{'}(l_{\textup{eq}})^{2}\right) \sin\theta_{\textup{eq}}(s)= 0 ,  \qquad s\in[0,l_{\textup{eq}}].
\end{equation}
The bending moment $M$ is proportional
to the rod's curvature, $M(s)=B \theta'(s)$, so that the equilibrium condition (\ref{elasticazzipost})
reveals the presence of an axial configurational force proportional to the square of the bending moment: $M(l_{\textup{eq}})^2/2B$, see
(Bigoni et al.~\cite{Bigoni-Bosi-DalCorso-Misseroni-2:2014}).

Restricting attention to the first buckling mode and exploiting symmetry, the investigation of equation~(\ref{elasticazzipost})
can be limited to the initial quarter part of the rod, $s\in[0, l_{\textup{eq}}/4]$, which may be treated
as the cantilever rod shown in the inselt of Fig.~\ref{variablelength}. The equilibrium configuration
can be expressed as the rotation field $\theta_{\textup{eq}}(s)$, solution of the following non-linear second-order differential problem
\begin{equation}
\label{elasticazzipostcr}
\begin{array}{lll}
\ds \theta_{\textup{eq}}^{''}(s) + \rho^2 \sin\theta_{\textup{eq}}(s)= 0 ,  \qquad s\in\left[0,\dfrac{l_{\textup{eq}}}{4}\right]  \\ [3 mm]
\theta_{\textup{eq}}(0) = 0,             \\ [3 mm]
\theta_{\textup{eq}}^{'}\left(\dfrac{l_{\textup{eq}}}{4}\right) =0,
\end{array}
\end{equation}
where the parameter $\rho$, representing the dimensionless axial thrust, has been introduced as
\begin{equation}\label{rho}
\rho=\sqrt{\dfrac{P}{B}+\dfrac{\theta_{\textup{eq}}^{'}(0)^2}{2}}.
\end{equation}

Integration of the elastica (\ref{elasticazzipostcr})$_1$ and the change of variable
\begin{equation}\label{cambiovarpost}
\upsilon=\sin\dfrac{\hat{\theta}}{2}, \qquad \upsilon\sin\phi(s)=\sin\dfrac{\theta_{\textup{eq}}(s)}{2},
\end{equation}
where $\theta_{\textup{eq}}\left(l_{\textup{eq}}/4\right)=\hat{\theta}$, yields the following differential problem
for the auxiliary field $\phi$
\begin{equation}
\label{systemequivalente}
\begin{array}{lll}
\ds \phi^{'}(s)= \rho \sqrt{1-\upsilon^2\sin^2\phi(s)} ,  \qquad s\in\left[0,\dfrac{l_{\textup{eq}}}{4}\right]  \\ [3 mm]
\phi(0) = 0,             \\ [3 mm]
\phi\left(\dfrac{l_{\textup{eq}}}{4}\right) =\dfrac{\pi}{2}.
\end{array}
\end{equation}
Further integration of (\ref{systemequivalente})$_1$ leads to the relation between the load parameter $\rho$
and the rotation measured at the free end of the cantilever $\hat{\theta}$ (which is an inflection point)\footnote{Would
the configurational force $M^2/2B$ be neglected, the following solution
is obtained
\begin{equation}
\label{mistpost}
\sqrt{\dfrac{P}{B}}\,l_{\textup{eq}}=4\mathcal{K}\left(\upsilon\right).
\end{equation}
}

\begin{equation}
\label{finalpostcritico}
\rho\,l_{\textup{eq}}=4\mathcal{K}\left(\upsilon\right),
\end{equation}
where $\mathcal{K}\left(\upsilon\right)$ is the complete elliptic integral of the first kind.

The configurational force, included in $\rho$, is a function of the curvature at the ends of the rod comprised between the two constraints,
namely $\theta_{\textup{eq}}^{'}(0)=\theta_{\textup{eq}}^{'}(l_{\textup{eq}})$, which can be obtained, through a multiplication of equation (\ref{elasticazzipostcr})$_1$ by $\theta_{\textup{eq}}^{'}$ and its integration, together with the boundary condition $\theta_{\textup{eq}}\left(l_{\textup{eq}}/4\right)=\hat{\theta}$, as
\begin{equation}\label{confforcepost}
\theta_{\textup{eq}}^{'}(0)=\sqrt{2\rho^2(1-\cos\hat{\theta})}=2\rho\upsilon,
\end{equation}
so that equation (\ref{rho}) may be rewritten as
\begin{equation}\label{caricofinalepostfede}
\dfrac{P}{B}=\rho^2\left(1-2\upsilon^2\right).
\end{equation}
The rotation field $\theta_{\textup{eq}}(s)$ can be expressed through the inversion of the relation (\ref{cambiovarpost})$_2$ as
\begin{equation}\label{tetapost}
 \theta_{\textup{eq}}(s) = 2 \textup{arcsin}\left(\upsilon\,\mbox{sn}\, (s \rho, \upsilon)\right),
\end{equation}
while the axial and transverse equations describing the shape of the elastica are obtained from integration of the following displacement fields
\begin{equation}
\label{campispost}
 x_1(s) = \int_0^s \cos\theta_{\textup{eq}}(\tau) \d \tau, \qquad x_2(s) = \int_0^s \sin\theta_{\textup{eq}}(\tau) \d \tau,
\end{equation}
as
\begin{equation}
\label{spostamentipost}
\begin{array}{ll}
x_1(s)= -s +\dfrac{2}{\rho} \left\{
\,\mbox{E}\left[\mbox{am}\left( s \rho , \upsilon\right), \upsilon\right]\right\},    \\ [4 mm]
x_2(s)= \dfrac{2\upsilon}{\rho}\left[1- \mbox{cn}(s\rho)\right],
\end{array}
\end{equation}
where the functions am, cn, and sn denote respectively the Jacobi amplitude, Jacobi cosine amplitude and Jacobi sine amplitude functions,
while $E(x,\upsilon)$ is the incomplete elliptic integral of the second kind of modulus $\upsilon$.
Note that equations (\ref{tetapost}) and (\ref{spostamentipost}) are valid for the entire structure, $s \in\left[0;l_{\textup{eq}}\right]$

Although the problem under consideration seems to be fully determined
by equations (\ref{finalpostcritico}), (\ref{tetapost}) and (\ref{spostamentipost}), the length $l_{\textup{eq}}$
is unknown because it changes after rod's buckling.
This difficulty can be bypassed taking advantage of symmetry, because the axial coordinate of the rod, for every unknown cantilever's length $l_{\textup{eq}}/4$, is $x_1(l_{\textup{eq}}/4)=L/4$, so that equation (\ref{spostamentipost})$_1$ gives
\begin{equation}\label{pincopalla}
-\dfrac{l_{\textup{eq}}}{4} +\dfrac{2}{\rho} \left\{\,\mbox{E}\left[\mbox{am}\left( \dfrac{l_{\textup{eq}}}{4} \rho , \upsilon\right), \upsilon\right]\right\}=\dfrac{L}{4},
\end{equation}
and therefore, inserting equation (\ref{finalpostcritico}) in equation (\ref{pincopalla}), it is possible to obtain the relation between the load parameter $\rho$ and the angle of rotation at the free edge of the cantilever $\hat{\theta}$ as
\begin{equation}\label{finalrhopostfede}
\rho=\dfrac{4}{L}\left\{2\,\mbox{E}\left[\mbox{am}\left( \mathcal{K}(\upsilon) , \upsilon\right), \upsilon\right]-\mathcal{K}(\upsilon)\right\}.
\end{equation}
The applied thrust $P$, normalized through division by the Eulerian critical load of the structure
$P_{\textup{cr}}=4\pi^2B/L^2$, is calculated from equation (\ref{caricofinalepostfede})
as a function of the kinematic parameter $\hat{\theta}$ through equation (\ref{cambiovarpost})$_1$ as
\begin{equation}
\label{sloffia}
\dfrac{P}{P_{\textup{cr}}}=\dfrac{4\,(1-2\upsilon^2)}{\pi^2}\left\{2\,\mbox{E}\left[\mbox{am}\left( \mathcal{K}(\upsilon) , \upsilon\right), \upsilon\right]-\mathcal{K}(\upsilon)\right\}^2.
\end{equation}

\begin{figure}[!htcb]
  \begin{center}
     \includegraphics[width= 12 cm]{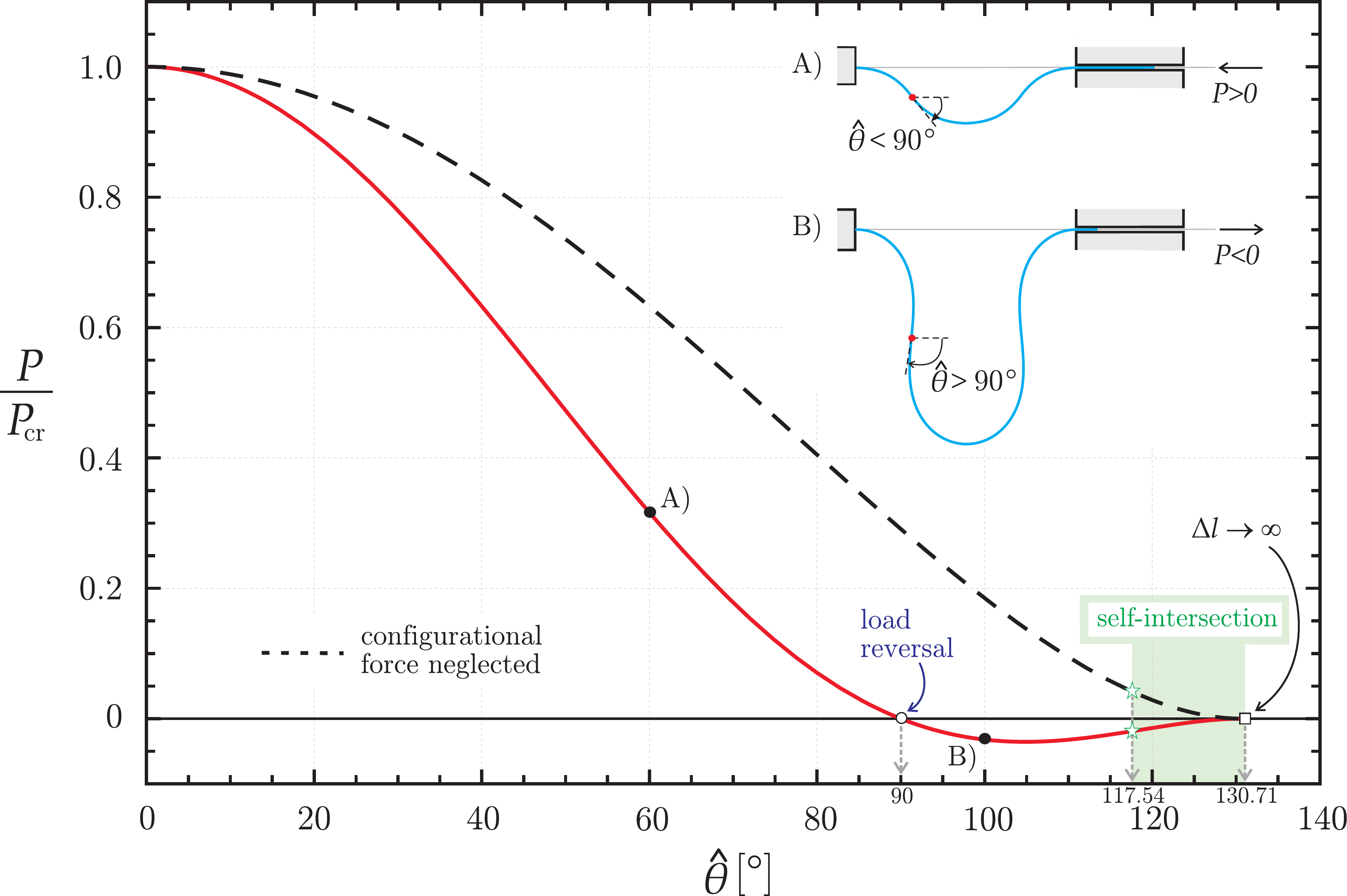}
\caption{\footnotesize Loading path of the structure sketched in Fig.~\ref{variablelength} (left),
expressed as the dimensionless applied dead load $P/P_{\textup{cr}}$ versus the rotation
 $\hat{\theta}$ at the inflexion point, equation (\ref{caricofinalepostfede}), shown with a red line.
The loading path
 is compared with that obtained by neglecting the configurational force (dashed black line),
 equation (\ref{mistpost}). Two deformed equilibrium configurations have been sketched in the inset showing that
  a compressive thrust (A) or tensile load (B) are necessary to guarantee
  equilibrium for $\hat{\theta}<90^\circ$ and $\hat{\theta}>90^\circ$, respectively.
For both solutions, after the condition  $\hat{\theta}\approx 117.54^\circ$ (marked with a green star) is met,  a
 self-intersecting elastica occurs, while load reversal
    does not occur when the configurational force is neglected. The rotation value $\hat{\theta}\approx 130.71^\circ$ is a limit condition
    corresponding to an infinite amount of injected rod length, $\Delta l\rightarrow\infty$,  and a null applied load, $P=0$.
}
\label{caricoangolopost}
  \end{center}
\end{figure}

Fig.~\ref{caricoangolopost} shows load P (divided by $P_{\textup{cr}}$) versus the rotation at the inflection point $\hat{\theta}$.
Also shown, for comparison, the dashed line representing the structural response when the configurational force
is neglected, equation ~(\ref{mistpost}).
It is noted that, although the critical load is not affected by the presence of the configurational force (because the bending moment is null before buckling),
the behaviour is strongly affected by it, so that the unstable postcritical path exhibits a \textit{force reversal}
 when $\hat{\theta}>90^\circ$ (confirmed also by experiments, see Sect.~\ref{exppost}),
absent when the configurational force is neglected.

According to equation (\ref{sloffia}), a self-equilibrated, but deformed, configuration ($P=0$) exists
for $\hat{\theta}=90^\circ$, where the rod is loaded only through the configurational force $M^2/2B$.

The length $\Delta l$, measuring the amount of rod \lq injected' through the sliding sleeve
\begin{equation}\label{additionalpost}
\Delta l=l_{\textup{eq}}-L,
\end{equation}
can be easily computed from equations (\ref{finalpostcritico}) and (\ref{finalrhopostfede}) in the following dimensionless form
\begin{equation}\label{deltallll}
\dfrac{\Delta l}{L}=\dfrac{\mathcal{K}(\upsilon)}{2\,\mbox{E}\left[\mbox{am}\left( \mathcal{K}(\upsilon) , \upsilon\right), \upsilon\right]-\mathcal{K}(\upsilon)}-1,
\end{equation}
whereas, considering the relationship (\ref{confforcepost}), the analytical expression for the \lq Eshelby-like' force (divided by $P_{\textup{cr}}$) becomes
\begin{equation}
\dfrac{M^2}{2BP_{\textup{cr}}}=\dfrac{8\upsilon^2}{\pi^2}\left\{2\,\mbox{E}\left[\mbox{am}\left( \mathcal{K}(\upsilon) , \upsilon\right), \upsilon\right]-\mathcal{K}(\upsilon)\right\}^2.
\end{equation}
When the rotation at the inflexion points exceeds $\hat{\theta}\approx 117.54^\circ$ (corresponding to $\Delta l\approx 5.61L$), the elastica self-intersects, so that the presented
solution holds true for rods capable of self-intersecting (as shown in \cite{Levyakov-Kuznetsov:2010, Bigoni-Libro:2012}).

Finally, it is noted that the condition
\begin{equation}
2\,\mbox{E}\left[\mbox{am}\left( \mathcal{K}(\upsilon) , \upsilon\right), \upsilon\right]-\mathcal{K}(\upsilon)=0
\end{equation}
occurs for $\hat{\theta}\approx 130.71^\circ$, representing a limit condition for which
the injected rod length becomes infinite, $\Delta l\rightarrow\infty$, and
the applied load becomes null, $P=0$, as evident
 from equations (\ref{deltallll}) and (\ref{sloffia}), respectively.

\section{Experimental}\label{exppost}

The structural system shown in Fig.~\ref{variablelength} (right) was designed and
manufactured at the Instabilities Lab (\href{http://ssmg.unitn.it/}{http://ssmg.unitn.it/}) of the University of Trento,
in such a way as to be loaded at prescribed displacement $\Delta l$,
with a continuous measure of the force $P$.

The displacement was imposed on the system with a loading machine MIDI 10 (from Messphysik).
During the test the applied axial force $P$ was measured using a MT1041 load cell (R.C. 500N) and the displacement $\Delta l$
by using the displacement transducer mounted on the testing machine. Data were acquired with a NI compactRIO system interfaced with Labview 2013 (from National Instruments).
The elastic rods employed during experiments were realized in solid polycarbonate strips (white 2099 Makrolon UV from Bayer, elastic modulus 2350 MPa), with dimensions 650 mm $\times$ 24 mm $\times$ 2.9 mm.
The sliding sleeve, 285 mm in length, was realized with 14 pairs of rollers from  Misumi Europe (Press-Fit Straight Type, 20 mm in diameter and 25 mm in length), modified to reduce friction.
The sliding sleeve was fixed to the two columns of the load frame, in the way shown in Fig.~\ref{variablelength} (right).

The influence of the curvature of the rollers on the amount of the configurational force generated at the end of the sliding sleeve was rigorously quantified in \cite{Bigoni-Bosi-DalCorso-Misseroni-2:2014} and
found to be negligible when compared to the perfect sliding sleeve model.

\begin{figure}[tp]
  \begin{center}
     \includegraphics[width= 16 cm]{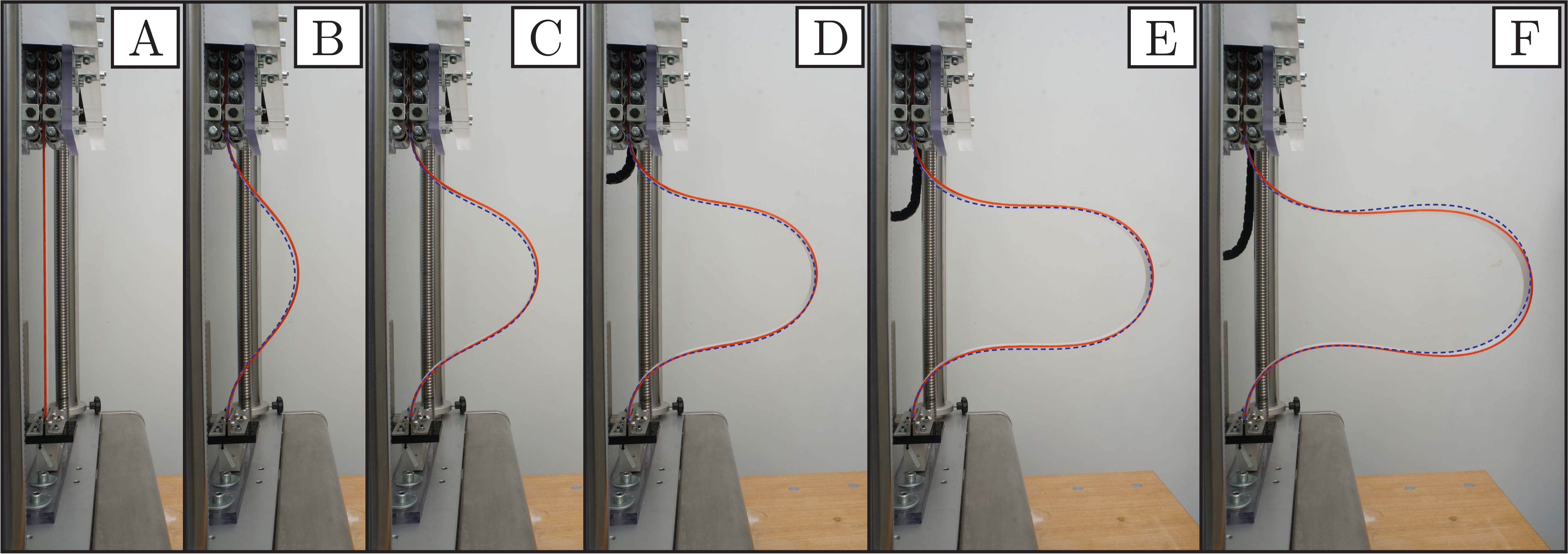}
\caption{\footnotesize A sequence of photos of experiments performed on the structure sketched in Fig.~\ref{variablelength} (right) in the
undeformed (A) and deformed configurations (B--F). The six photos show the displacement-controlled experiment at
$\Delta l=\left\{0;0.14;0.41;0.82;1.21;1.58\right\}L$, corresponding to measured
loads $P=\left\{0;0.62;0.26;0.06;0;-0.02\right\}P_{\textup{cr}}$. The calculated postbuckling deformed configurations
are superimposed to the photos as dashed lines.
}
\label{sequenzepost}
  \end{center}
\end{figure}

A sequence of six photos taken during the experiment with $L=360$ mm is reported in Fig.~\ref{sequenzepost}, together with the theoretical elastica, which are found to be nicely superimposed to the
experimental deformations.

\begin{figure}[tp]
  \begin{center}
     \includegraphics[width= 16 cm]{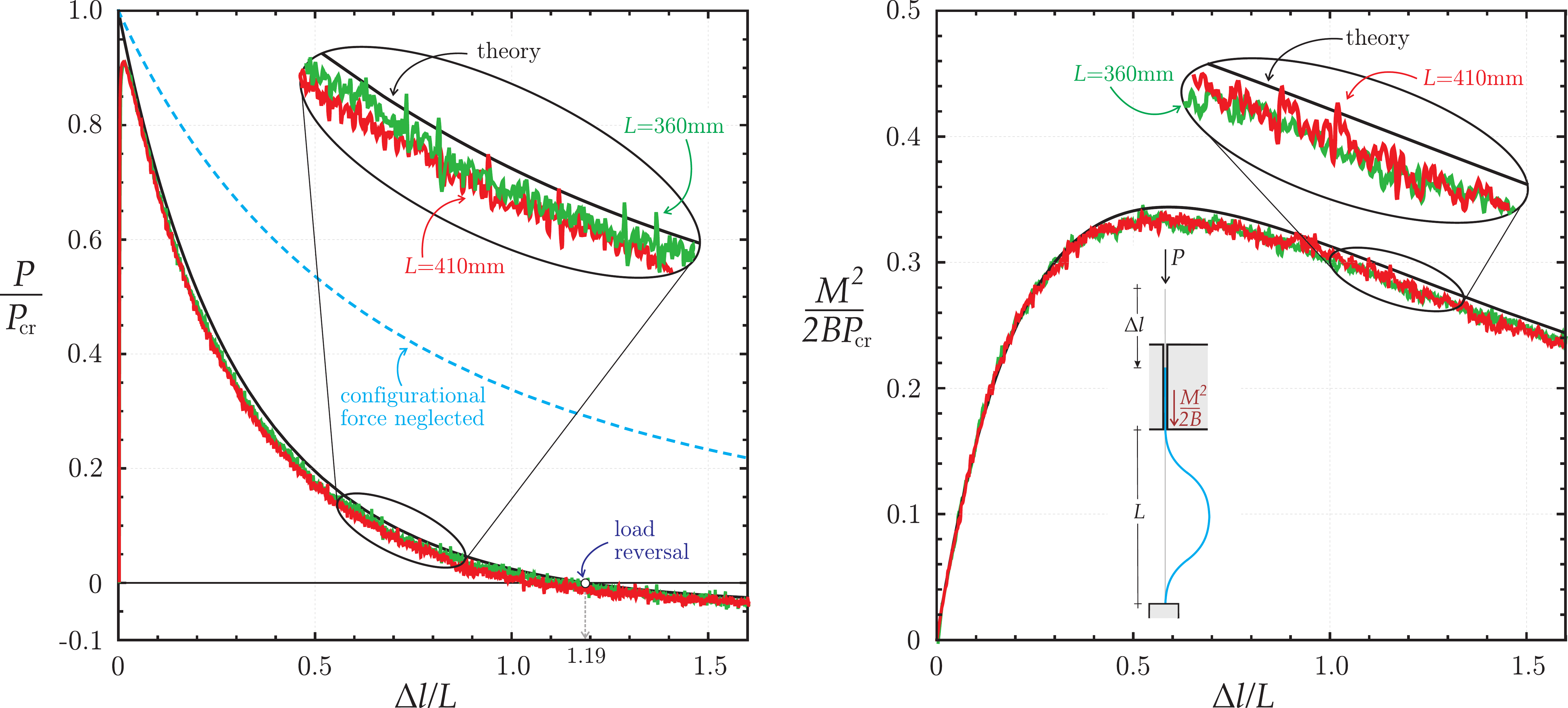}
\caption{\footnotesize Comparison between theoretical predictions (black curve) and experimental
results for two different constraint distances $L=360$ mm (green curve) and $L=410$ mm (red curve).
Dimensionless axial thrust $P/P_{\textup{cr}}$ (left) and dimensionless configurational
force $M^2/(2BP_{\textup{cr}})$ (right) are plotted versus
the controlled dimensionless length $\Delta l/L$ (measuring the amount of rod \lq injected' into the sliding sleeve).
The postcritical path obtained neglecting the configurational force is reported as a blue dashed curve in the graph on the left.
}
\label{esperimentipost}
  \end{center}
\end{figure}

Experimental results are reported in Fig.~\ref{esperimentipost} and compared with the presented theoretical solution and the solution obtained by
neglecting the configurational force (dashed line). While the former solution is in excellent agreement with experimental
results performed for two different distances between the two constraints ($L=360$ mm and $L=410$ mm),
the latter solution reveals that neglecting the configurational force introduces a serious error.
Note that two different lengths $L$ have been tested only to verify the robustness of the model, because size effects are theoretically not found with the assumed parametrization, which makes results independent also on the bending stiffness $B$.

The dimensionless applied load $P/P_{\textup{cr}}$ versus dimensionless length $\Delta l/L$ (Fig.~\ref{esperimentipost}, left), measuring the amount of elastic rod \lq injected' into the sliding sleeve,
confirms a load reversal at $\Delta l/L\approx 1.19$.
We note that the configurational force does not influence the qualitative shape of the elastica, but the amount of deflection, so that neglecting it leads to
a strong and completely unacceptable estimation of the amplitudes.

Finally, the dimensionless configurational force $M^2/2BP_{\textup{cr}}$ reported as a function of $\Delta l/L$ (Fig.~\ref{esperimentipost} on the right)
shows its strong influence on postcritical behaviour,
such that this force grows to make up one third of the rod's critical load.

\section{Conclusion}

Instabilities occurring during injection of an elastic rod through a sliding sleeve were investigated, showing the presence of a strong
configurational force.
This force causes a force reversal during a softening post-critical response and
has been theoretically determined and experimentally validated, so that it is now ready for exploitation in the design of compliant mechanisms.
Moreover, configurational forces as investigated in this paper can also be generated through growth or swelling, so that the can play
an imortant role in the description of deformation processes of soft matter.

\vspace*{5mm} \noindent
{\sl Acknowledgments } Financial support from the ERC Advanced Grant \lq Instabilities and nonlocal multiscale modelling of materials'
FP7-PEOPLE-IDEAS-ERC-2013-AdG (2014-2019) is gratefully acknowledged.


\end{document}